\documentclass[a4paper,11pt]{article}
\usepackage{pos}

\title{iDMEu: An initiative for Dark Matter in Europe and beyond}

\author[a]{Marco CIRELLI}
\author[b]{Caterina DOGLIONI}
\author[c]{Federica PETRICCA}

\affiliation[a]{Laboratoire de Physique Th\'eorique et Hautes \'Energies (LPTHE), CNRS \& Sorbonne Universit\'e, \\
   4 Place Jussieu, Paris, France}

\affiliation[b]{School of Physics and Astronomy, University of Manchester,\\
M13 9PL Manchester, United Kingdom}

\affiliation[c]{Max-Planck-Institut f\"ur Physik, \\
D-80805 M\"unchen, Germany}

\emailAdd{iDMEu-jenaa-eoi-organizers@cern.ch}
\emailAdd{marco.cirelli@gmail.com}
\emailAdd{caterina.doglioni@cern.ch}
\emailAdd{petricca@mpp.mpg.de}

\abstract{We introduce the {\em initiative for Dark Matter in Europe and beyond (iDMEu)}, a collective effort by a group of particle and astroparticle physicists to set up an online resource meta-repository, a common discussion platform and a series of meetings on everything concerning Dark Matter. This document serves as a status report as well as a citable item concerning iDMEu.} 

\FullConference{XVIII International Conference on Topics in Astroparticle and Underground Physics (TAUP2023)\\
 28.08 $-$ 01.09.2023\\
University of Vienna\\}

\begin{document}
\maketitle

\section{Introduction and rationale}
Understanding dark matter (DM), how it was produced in the early universe, what its nature is and where it is located in the cosmos is one of the fundamental physics problems of our century.
The community working on DM is active and very diverse, including particle physics theorists, cosmologists and astrophysicists with a wide range of interests, as well as particle physics experimentalists focusing on collider, fixed-target, beam-dump, direct and indirect DM detection experiments, as well as dedicated axion/ALP (axion-like particles) experiments.
Given the diversity of the DM community, discovering or constraining dark matter requires broad discussions.

\smallskip

At the \href{https://jenas-2019.ijclab.in2p3.fr/}{JENAS workshop} held in Orsay (France) in October 2019, jointly organized by \href{https://ecfa.web.cern.ch/}{ECFA} (the European Committee for Future Accelerators), \href{https://www.nupecc.org/}{NuPECC} (the Nuclear Physics European Collaboration Committee) and \href{https://www.appec.org/}{APPEC} (the Astroparticle Physics European Consortium), a call for Expressions of Interest was launched, inviting projects with interest spanning the high energy physics, astroparticle physics and nuclear physics community.

A small number of colleagues working on DM (the {\bf iDMEu} proponents) decided to submit such a \href{https://indico.cern.ch/event/869195/}{Expression of Interest} for an {\em initiative on Dark Matter in Europe and beyond} ({\bf iDMEu}) and they collected endorsements from the community. 
In the past few years, {\bf iDMEu} has equipped itself with a light management structure, has participated in some events (notably the \href{https://jenas-2019.ijclab.in2p3.fr/}{2nd JENAS-Seminar} in Madrid in 2022), has organized some events and has worked on its missions. 

The present document serves as a summary of the characteristics and status of {\bf iDMEu} and as a citable permanent resource for the colleagues using the services and facilities that {\bf iDMEu} provides.

\subsection{iDMEu goals}
 iDMEu enables to exploit synergies and to highlight the complementarity of different DM communities by: \vspace{-0.3cm}
\begin{itemize}
\setlength\itemsep{0em}
\item[$\triangleright$] collecting and classifying DM resources,
\item[$\triangleright$] developing a common platform to facilitate (and participate in) new cross-community scientific collaborations,
\item[$\triangleright$] helping communicate a common DM story for different audiences.
\end{itemize}
Concretely, that means:
\vspace{-0.3cm}
\begin{itemize}
\setlength\itemsep{0em}
\item[$\diamond$] creating an online meta-repository of DM resources (see section \ref{sec:website}),
\item[$\diamond$] organizing regular `town-hall' meetings (see section \ref{sec:events}), typically once or twice a year depending on the response of the community. The focus will be on: i) the challenges faced by each community and the way ahead, ii) the cross-communication among the communities. 
\end{itemize}\vspace{-0.2cm}
It is important to stress that {\bf iDMEu} does not intend to replace or reproduce existing similar efforts, but rather complement them and act as a hub/aggregator. 

\section{The iDMEu structure}
The colleagues working on {\bf iDMEu} are organized in the following different groups.\vspace{-0.3cm}
\begin{itemize}
\setlength\itemsep{0em}
\item The {\bf iDMEu Proponents} are the signatories of the initial \href{https://indico.cern.ch/event/869195/}{Expression of Interest}. They identified the need for closer connections among the different DM communities and took the initiative of funding {\bf iDMEu}. They managed the {\bf iDMEu}'s activities during the first two years. The group consisted of M.~ Cirelli (CNRS and Sorbonne University, Paris, France), C.~Doglioni (University of Manchester, UK), F.~Petricca (Max-Planck-Institut f\"ur Physik, M\"unchen, Germany), F.~Reindl (HEPHY and TU Wien, Austria), G.~Lanfranchi (LNF-INFN, Italy), J.~Monroe (Royal Holloway London, UK), S.~Pascoli (IPPP Durham, UK \& Bologna University, Italy) and E.~Cuoco (EGO European Gravitational Observatory \& Scuola Normale Superiore, Pisa, Italy).
\item The {\bf iDMEu Organizers} actively manage the initiative and steer it towards the achievement of its goals. As of September 2023, the group consists of M.~ Cirelli, C.~Doglioni, F.~Petricca (the authors of this document) and G.~Zaharija\v s (University of Nova Gorica, Slovenia). This group is bound to be extended, in order to include colleagues who share the goals and come from DM sub-communities that are currently under-represented (e.g.~cosmology and astrophysics). 
\item The {\bf iDMEu Curators} work in creating the {\bf iDMEu} resource content. Under the supervision of one or more Organizers, they collect, classify and elaborate the information, and develop the website. 
The group consists of Early Career Researchers including Bachelor's, Master's, PhD students as well as postdocs from various institutions in Europe. They often engage in these activities as part of their Bachelor's, Master's, or Ph.D. research projects or theses.. 
This allowed them to efficiently obtain a wide view of the DM research panorama.
\end{itemize}
The {\bf iDMEu} Organizers regularly report to the sponsoring agencies ECFA, NuPECC and APEC, from which they have received a small amount of funding\footnote{The bulk of the funding, for the website and trips, comes however from the Organizers' individual research funds.}. In 2022 the agencies have set up a Task Force: a group of external colleagues that met a few times online to provide feedback to {\bf iDMEu}.

\section{The iDMEu online meta-repository and website} \label{sec:website}
\begin{figure}[!t]
\center
\includegraphics[width=0.9\textwidth]{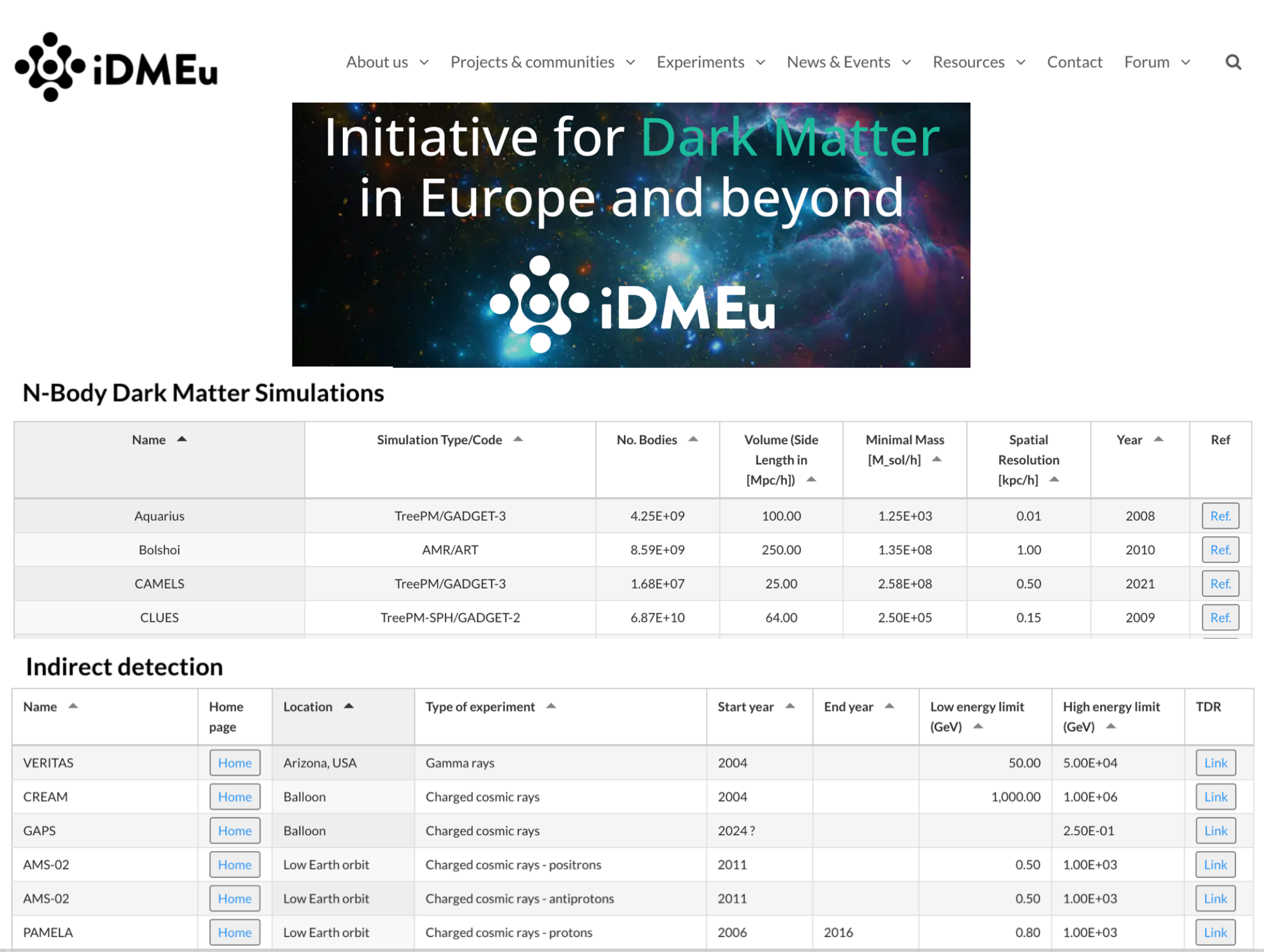}
\caption{\label{figure}Excerpts from the {\bf iDMEu} online meta-repository: logo, menu and background image (top), excerpt of the tables on Indirect DM Detection experiments and N-body numerical simulations (bottom).}
\end{figure}

The meta-repository is accessible under the permanent URL \href{http://www.idmeu.org}{http://www.idmeu.org}. Beside a general description of {\bf iDMEu} and a list of DM-related projects and events, it features the following.
\vspace{-0.5cm}
\begin{itemize}
\setlength\itemsep{0em}
\item A collection of tables, listing many DM-related `experiments' (intended in a broad sense, as the term includes numerical simulations and astrophysical observations) and their characteristics, with links to the homepages and to the main technical publications. These tables can be searched, ranked and filtered. They represent the major effort deployed by the {\bf iDMEu} Organizers and Curators since the onset of the initiative. They are intended to be upgraded regularly, e.g.~on a monthly basis.
\item Another collection of tables, listing many DM-related resources such as reviews, lectures, outreach items (books, articles, videos...). 
\item A moderated discussion forum put at disposal of the community and organized around threads. 
\end{itemize}
An excerpt of the website is provided in fig.~\ref{figure}. The website is intended as in continuous evolution. A feedback form is offered. New Curators will be hired to maintain and extend the resources.  

\section{The iDMEu events} \label{sec:events}
{\bf iDMEu} held its \href{https://indico.cern.ch/event/1016060/}{kick-off meeting} on 10-12 May 2021, as a fully online event. About 375 participants registered and a large fraction attended at least a part of the meeting. The program was organized as follows:
\vspace{-0.3cm}
\begin{itemize}
\setlength\itemsep{0em}
\item About 15 overview talks by invited experts of each sub-community (e.g.~`DM DD', `axions', `theory' etcetera), highlighting the challenges of their field and the needed input. 
\item Breakout discussion sessions among the participants.
\item An outreach session featuring several examples of successful outreach projects, for different audiences and in diverse formats.
\item A Q\&A session: DM-related basic questions were collected anonymously ahead of the event and two experts were asked to provide a pedagogic mini-course answering them.
\end{itemize}

{\bf iDMEu} held \href{https://indico.cern.ch/event/1309317/}{1st `town hall' meeting} on 1 September 2023, as a satellite event of the XVIII TAUP conference (these proceedings). About 100 participants registered, about half of them attended in person in Vienna and some others online. A preliminary version of the online meta-repository and the discussion forum was presented. The issues tackled by the contributions included: the DM theory landscape, the potential synergies in detector technologies for dark matter experiments and the efforts on DM complementarity within the \href{https://snowmass21.org/}{Snowmass process}.

{\bf Acknowledgements:} The authors warmly thank all the iDMEu Proponents and the Curators (see text) for their work on iDMEu, as well as the sponsoring agencies: \href{https://ecfa.web.cern.ch/}{ECFA} (the European Committee for Future Accelerators), \href{https://www.nupecc.org/}{NuPECC} (the Nuclear Physics European Collaboration Committee) and \href{https://www.appec.org/}{APPEC} (the Astroparticle Physics European Consortium).

%

\end{document}